\documentclass[12pt]{iopart}

\usepackage{graphicx}

\begin{document}

\title{Floquet exceptional points and chirality in non-Hermitian Hamiltonians}

\author{Stefano Longhi}

\address{Dipartimento di Fisica,
Politecnico di Milano, Piazza L. da Vinci 32, I-20133 Milano, Italy}
\ead{longhi@fisi.polimi.it}
\begin{abstract}
Floquet exceptional points correspond to the coalescence of two (or more) quasi-energies and corresponding Floquet eigenstates of a time-periodic non-Hermitian Hamiltonian. They generally arise when the oscillation frequency satisfies a multi-photon resonance condition. Here we discuss the interplay between Floquet exceptional points and the chiral dynamics observed, over several oscillation cycles, in a wide class of non-Hermitian systems when they are slowly cycled in opposite directions of parameter space. 
\end{abstract}

\vspace{2pc}

\maketitle

\section{Introduction}

Non-Hermitian Hamiltonians are used in both classical and quantum physics to 
provide an effective description of the behavior of open systems \cite{r1,r2,r3,r4,r5}. Examples of effective non-Hermitian Hamiltonians
are found in nuclear physics, atomic and molecular physics, atom optics, laser physics, microwave and optical systems, acoustics, mechanical and electronic circuits, etc. 
Interesting physical effects arise at so-called non-Hermitian degeneracies or exceptional points (EPs) \cite{r6,r7,r8,r9,r10}, 
at which two (or more) eigenvalues and corresponding eigenvectors of the underlying non-Hermitian operator coalesce. 
EPs arise, for example, in $\mathcal{PT}$-symmetric systems \cite{r11,r12}, i.e. systems described by a non-Hermitian Hamiltonian which is invariant under
the combined action of parity inversion and time reversal, where simultaneous coalescence of energies and eigenstates is generally found at the symmetry breaking point, i.e. the transition point in parameter space that separates the energy spectrum of the Hamiltonian from being entirely real to become complex. EPs are at the heart of many intriguing physical phenomena appearing in several systems that experience gain or loss. They have been predicted and observed in a wide variety of physical systems, including atomic and molecular systems \cite{r13,r14,r15}, microwave cavities and waveguides \cite{r16,r17,r18}, electronic circuits \cite{r19}, optical structures \cite{r20,r21}, Bose-Einstein condensates \cite{r22,r22bis}, acoustic cavities \cite{r23}, non-Hermitian Bose-Hubbard models \cite{r24}, exciton-polariton billiards \cite{r24bis}, opto-mechanical systems \cite{r25} and many others.  Besides of their theoretical interest, EPs can find important applications, for example in the design and realization of unidirectionally invisible media \cite{r26,r27,r27bis,r28,r28bis}, for asymmetric mode switching \cite{r18,r29} and topological energy transport \cite{r25},  for the design of novel laser devices \cite{r30,r31,r32,r33,r34}, for optical sensing \cite{r35} and polarization mode conversion \cite{r36}.  The dynamical properties associated to the encircling of an EP and the chirality of EPs arising from breakdown of the adiabatic theorem  have received a great attention in recent years \cite{
r16,r17,r18,r25,r36,r37,r38,r39,r40,r41}. A remarkable property observed when encircling an EP is a topological-robust  state-flip for quasi-static cycling \cite{r15,r17,r42}, and the chiral behavior associated to breakdown of adiabaticity for dynamical circling \cite{r18,r25,r36,r37,r38,r40}. In the former case 
state flip arises from to the branch point character of the degeneracy that causes a gradual transition between the intersecting complex Riemann sheets.
In the latter case, the system dynamically evolves around an EP and non-adiabatic transitions prevent state flip when the EP is encircled in one circulation direction \cite{r37,r38,r40}, while in the opposite direction the adiabatic evolution is kept.  As a result, a chiral behavior is obtained, i.e. a different final state is selected when dynamically encircling an EP in clockwise or counter-clockwise directions. Recent experiments \cite{r18,r25} demonstrated  chirality of EP cycling, originally predicted in Refs.\cite{r37,r38}, and raised a great interest owing to potential applications of EPs to topological energy transport \cite{r25}, asymmetric mode switching \cite{r18,r29} and polarization control of light \cite{r36}. Asymmetric non-adiabatic transitions observed when the EP is dynamically encircled are deeply rooted in the non-Hermitian nature of the underlying dynamics and have been explained in terms of the Stokes phenomenon of asymptotics \cite{r38}, the stability loss delay phenomenon \cite{r40}, and the asymmetric nature of transitions induced by non-Hermitian perturbations \cite{r43}.\\  
In this work we disclose a different and rather general mechanism of chirality in a non-Hermitian Hamiltonian which undergoes repeated cycles in parameter space. Rather than arising from encircling a static EP, chirality is here rooted into the appearance of Floquet EPs. A Floquet EP corresponds to the coalescence of two (or more) quasi-energies and corresponding Floquet eigenstates of a time-periodic non-Hermitian Hamiltonian. 
When the quasi-energy spectrum of the non-Hermitian Hamiltonian is real, Floquet EPs generally arise when the frequency $\omega$  of cycling is tuned to a multi-photon resonance of quasi-energies. In this case, a chiral dynamics is observed when the system is periodically and slowly cycled in opposite directions of parameter space, i.e. 
 a different final state is selected in the two directions of rotation starting from the same initial state.
 \section{Floquet exceptional points}
 Let us consider a finite-dimensional and time-periodic Hamiltonian, which is defined by a $N \times N$ non-Hermitian matrix $\mathcal{H}=\mathcal{H}(t)$ with time-periodic coefficients, $\mathcal{H}(t+T)=\mathcal{H}(t)$ with $T= 2 \pi / \omega$. The dynamical equations of motion of the vector amplitudes $\mathbf{a}(t)=(a_1(t), a_2(t),...,a_N(t))^T$ read
 \begin{equation}
 i \frac{d \mathbf{a}}{dt}= \mathcal{H}(t) \mathbf{a}.
 \end{equation}
 According to Floquet` s theory, the solution to Eq.(1) with assigned initial condition $\mathbf{a}(0)$ is given by (see, for instance, \cite{r44})
 \begin{equation}
 \mathbf{a}(t)= \Phi(t) \exp(-i \mathcal{R}t) \mathbf{a}(0)
 \end{equation}
 where $\Phi(t+T)=\Phi(t)$ is a $N \times N$ periodic matrix with $\Phi(0)=1$ (identity matrix) and $\mathcal{R}$ is a time-independent $N \times N$ matrix whose eigenvalues $\mu_l$ ($l=1,2,...,N$) are the quasi-energies (Floquet exponents). Quasi-energies are defined apart from integer multiples of $\omega$, and therefore two quasi-energies are degenerate if they are equal or differ each other by multiples of $\omega$. For the sake of definiteness, we will take the real parts of quasi-energies in the range $-\omega/ 2 \leq {\rm Re} (\mu_l) \leq \omega /2$. Indicating by $\mathbf{q}_n$ the eigenvectors of $\mathcal{R}$ with corresponding eigenvalues $\mu_n$, i.e. $\mathcal{R} \mathbf{q}_n= \mu_n \mathbf{q}_n$ ($n=1,2,...,N$), the $N$ Floquet eigenstates of Eq.(1) are obtained from Eq.(2) by letting $\mathbf{a}(0)=\mathbf{q}_n$ and read explicitly 
  \begin{equation}
 \mathbf{f}^{(n)}(t)=\Phi(t) \mathbf{q}_n \exp(-i \mu_n t).
 \end{equation}
 Note that a Floquet eigenstate satisfies the condition $ \mathbf{f}^{(n)}(t+T)= \mathbf{f}^{(n)}(t) \exp(-i \mu_n T)$. It is worth observing that the dynamics of Eq.(1), mapped at discretized times $t=0,T,2T,3T,...$, is equivalent to the one of a time-independent system with Hamiltonian $\mathcal{R}$. In fact, for $t=nT$ ($n=0,1,2,...$), from Eq.(2) one has $\mathbf{a}(t)=\exp(-i \mathcal{R}t) \mathbf{a}(0)$. Since $\exp(-i \mathcal{R}t)$ can be viewed as the propagator of the time-independent Hamiltonian $\mathcal{R}$,  at $t=nT$ one has $\mathbf{a}(t)=\mathbf{v}(t)$ with
 \begin{equation}
 i \frac{d \mathbf{v}}{dt}=\mathcal{R} \mathbf{v}
 \end{equation} 
 and $ \mathbf{v}(0)= \mathbf{a}(0)$. For a Hermitian system, the quasi-energies are real and the Floquet eigenstates are linearly independent functions, even in presence of quasi-energy degeneracy (i.e. coalescence of two or more quasi-energies).  Therefore for a Hermitian Hamiltonian $\mathcal{H} (t)$ the most general solution to Eq.(1) can be written as a superposition of Floquet eigenstates
 \begin{equation}
 \mathbf{a}(t)=\sum_{n=1}^N \alpha_n \mathbf{f}^{(n)}(t)
  \end{equation}
 with coefficients $\alpha_n$ determined by the initial condition. However, for a non-Hermitian system ($\mathcal{H}^{\dag} \neq \mathcal{H}$) two (or more) quasi-energies and corresponding Floquet eigenstates can simultaneously coalesce, corresponding to the appearance of a Floquet EP.  For a Floquet EP of order $M$, with $2 \leq M \leq N$, one can assume $\mu_{y_1}=\mu_{y_2}=...=\mu_{y_M}$ and $\mathbf{f}^{(y_1)}(t)=\mathbf{f}^{(y_2)}(t)=...=\mathbf{f}^{(y_M)}(t)$  for a set of $M$ distinct indices $y_1,y_2,...,y_M$, while the remaining $(N-M)$ Floquet states $\mathbf{f}^{(n)}(t)$ ($n \neq y_1, y_2,...,y_M$) are linearly independent functions and distinct from $\mathbf{f}^{(y_1)}(t)$.  Clearly,  at a Floquet EP of order $M$ the eigenvalue $\mu_{y_1}=\mu_{y_2}=...=\mu_{y_M}$ of the matrix $\mathcal{R}$ is a defective eigenvalue with a geometric multiplicity smaller than its algebraic multiplicity and with $\mathbf{q}_{y_1}=\mathbf{q}_{y_2}=...=\mathbf{q}_{y_M}$. In this case, one can introduce $M$ generalized eigenvectors $\mathbf{Q}_n$ of $\mathcal{R}$, with $n=y_1,y_2,..,y_M$, defined as $\mathbf{Q}_{y_1}=\mathbf{q}_{y_1}$ and $(\mathcal{R}-\mu_{y_1})\mathbf{Q}_{y_n}=\mathbf{Q}_{y_{n-1}}$ for $n=2,3,..,M$, or by similar relations obtained by some permutation of indices $y_1$, $y_2$, .., $y_M$. The set of generalized eigenvectors $\mathbf{Q}_{y_n}$ are linearly independent and, together with the remaining eigenvectors $\mathbf{q}_n$ ($n \neq y_1,y_2,...,y_M$), form a complete basis. Likewise, for the defective Floquet quasi-energy $\mu_{y_1}$ one can introduce a set of $M$ linearly-independent {\it generalized} Floquet eigenstates $\mathbf{F}^{(y_n)}(t)$, defined by 
 \begin{equation}
 \mathbf{F}^{(y_n)}(t)= \Phi(t) \left( \sum_{k=0}^{n-1} \beta_{n-1,k} t^{n-k-1} \mathbf{Q}_{y_{k+1}} \right) \exp(-i \mu_{y_1} t)
 \end{equation} 
  ($n=1,2,...,M$), where the coefficients $\beta_{n,k}$ are given by
  \begin{equation}
  \beta_{n,k}=i^k \frac{n ! }{(n-k)!}
  \end{equation}
($k=0,1,...,n$). Note that $\mathbf{F}^{(y_1)}(t)=\mathbf{f}^{(y_1)}(t)$. For a non-Hermitian system with a Floquet EP, the most general solution to Eq.(1) is given by [compare with Eq.(5)]
\begin{equation}
\mathbf{a}(t)=\sum_{n=y_1,y_2,...,y_M}^M \alpha_n \mathbf{F}^{(n)}(t)+\sum_{n \neq y_1,y_2,...,y_M} \alpha_n \mathbf{f}^{(n)}(t)
\end{equation}
 with coefficients $\alpha_n$ determined by the initial condition.\\
 In the following analysis, we will focus our attention to a time-periodic Hamiltonian with an entirely real quasi-energy spectrum. In this limiting case, it is worth considering the long-time behavior of the solution to Eq.(1). If all the $N$ Floquet eigenstates $\mathbf{f}^{(n)}(t)$ are linearly independent, i.e. in the absence of Floquet EPs, according to Eqs.(3) and (5) it follows that the solution to Eq.(1) remains bounded as $t \rightarrow \infty$. Conversely, if the time-periodic Hamiltonian $\mathcal{H}$ has a Floquet EP of order $M$, according to Eqs.(6) and (8) the solution $\mathbf{a}(t)$ generally shows  an algebraic secular growth in time with the asymptotic behavior $\mathbf{a}(t) \sim t^s \Phi(t) \mathbf{q}_{y_1} \exp(-i \mu_{y_1} t)=t^s \mathbf{f}^{(y_1)}(t)$ with exponent $s \leq M-1$, indicating that the degenerate Floquet eigenstate $\mathbf{f}^{(y_1)}(t)$ is the dominant state of the dynamics. Note that the secular growth is prevented only when $\alpha_n=0$ for $n=y_2,..,y_M$ in Eq.(8), i.e. when the initial state $\mathbf{a}(0)$ has no projection into the generalized eigenstates $\mathbf{Q}_{y_2}$, $\mathbf{Q}_{y_3}$, .., $\mathbf{Q}_{y_M}$. An important and immediate consequence of the dominance of the Floquet eigenstate $\mathbf{f}^{(1)}(t)$ is the breakdown of the adiabatic theorem induced by a Floquet EP for a slowly-cycled Hamiltonian. In fact, let us assume an arbitrarily small oscillation period $T \rightarrow 0$ and let us prepare the initial state $\mathbf{a}(0)$ in an instantaneous eigenvector of $\mathcal{H}(0)$. Even though the instantaneous eigenvalues $\lambda_n(t)$ of $\mathcal{H}(t)$ are distinct each other and separated by some finite gap  in the entire oscillation cycle, owing to the existence of the EP at long times the state $\mathbf{a}(t)$ becomes $\mathbf{f}^{(1)}(t)$ dominated, indicating that the dynamics ceases to be adiabatic after many oscillation cycles. This point will be clarified in subsequent Secs.3.2 and 3.3 and exemplified in Sec.4.

\section{Nonadiabatic transitions and chirality induced by Floquet exceptional points}
\subsection{Model}
 Let us consider a class of non-Hermitian time-periodic Hamiltonians $\mathcal{H}(t)$ of the form
\begin{equation}
\mathcal{H}(t)=\mathcal{H}_0+\sum_k R_k(t) \mathcal{H}_k
\end{equation}
where $\mathcal{H}_0$ and $\mathcal{H}_k$ are time-independent $N \times N$ matrices and $R_k(t)$ are the elements of a complex parameter vector $\mathbf{R}=\mathbf{R}(t)$ of arbitrary dimension that is slowly cycled at frequency $\omega= 2 \pi /T$. The parameter vector $\mathbf{R}$  is assumed to have zero mean and composed by positive-frequency components solely, i.e. $\mathbf{R}(t)= \sum_{n=1}^{\infty} \mathbf{R}^{(n)} \exp(i n \omega t)$. After the time $T= 2 \pi/\omega$ from the initial time $t=0$, the vector $\mathbf{R}(t)$ has described a closed loop $\mathcal{C}$ in multivariable parameter space. For $\omega>0$ we say that the loop $\mathcal{C}$ is circulated 'clockwise' in parameter space. By reversing the sign of $\omega$, i.e. by considering the dynamical behavior of the system backward in time, we say that the loop $\mathcal{C}$ is circulated 'counter-clockwise'. The main result of the present work is to show that a chirality in the system arises in the presence of a Floquet EP.

\subsection{Quasi-energy spectrum, Floquet eigenstates and Floquet EPs}
We can prove the following two theorems.\\
\\
{\bf Theorem 1.} Let $\mathcal{H}(t)$ be a periodic Hamiltonian of the form (9) with period $T= 2 \pi / \omega$ and let us assume that:\\
(i) The eigenvalues $\lambda_n$ of $\mathcal{H}_0$ are real and distinct, with $\lambda_1 < \lambda_2<...< \lambda_N$.\\
(ii) For any arbitrary two eigenvalues $\lambda_n$ and $\lambda_m$ of $\mathcal{H}_0$, the difference $|\lambda_n-\lambda_n|$ is not a multiple of $\omega$.\\
Then the quasi-energy spectrum of $\mathcal{H}(t)$ is entirely real, the quasi-energies are distinct and given by $\mu_n=\tilde{\lambda}_n$, where $\tilde{\lambda}_n=\lambda_n-s \omega$ and $s$ is an integer such that $ - \omega/2 \leq \tilde{\lambda}_n < \omega/2$. \footnote{We say that $\tilde{\lambda}_n$ is the value $\lambda_n$ folded inside the Floquet interval $(-\omega/2, \omega/2)$ of quasi-energies.}\\
{\it Proof. } After setting $\mathcal{S}(t)=\sum_k R_k(t) \mathcal{H}_k$, let us note that, since $R_k(t)$ are composed by positive-frequency components solely (for a clockwise circulation of the loop $\mathcal{C}$, $\omega>0$) or by  negative-frequency components solely (for a counter-clockwise circulation of the loop $\mathcal{C}$, $\omega<0$), we can expand $\mathcal{S}(t)$ in Fourier series as 
\begin{equation}
\mathcal{S}(t)=\sum_{k=1}^{ \infty} \mathcal{S}^{(k)} \exp(ik \omega t).
\end{equation}
 We then look for a solution to Eq.(1) of the Floquet form, i.e. of the form
\begin{equation}
\mathbf{a}(t)=\mathbf{f}(t)=\exp (- i \mu t) \sum_{l=-\infty}^{\infty} \mathbf{a}^{(l)} \exp(i \omega l t)
\end{equation}   
where $\mu$ is the quasi-energy. Substitution of Anstaz (11) into Eq.(1) and using Eq.(10), after equating the terms oscillating like $\sim \exp(i \omega l t)$  one readily obtains 
\begin{equation}
\left( \mu-l \omega- \mathcal{H}_0 \right) \mathbf{a}^{(l)}=\sum_{k=1}^{\infty} \mathcal{S}^{(k)} \mathbf{a}^{(l-k)}.
\end{equation}
Note that, since the sum on the right hand side of Eq.(12) runs for positive integers $k$ solely (this is because of the one-sided Fourier spectrum of $\mathbf{R}(t)$), the vector $\mathbf{a}^{(l)}$ depends solely on the vectors  $\mathbf{a}^{(h)}$ with index $h<l$. Therefore, we can find $N$ distinct solutions to Eq.(12) by assuming $\mu=\lambda_n$ ($n=1,2,....,N$) and, correspondingly:
\begin{equation}
\mathbf{a}^{(l)}= \left\{
\begin{array}{cc}
0 & l <0 \\
\mathbf{w}^{(n)} & l=0 \\
 \left( \lambda_n - l \omega - \mathcal{H}_0 \right)^{-1} \sum_{k=1}^{l} \mathcal{S}^{(k)} \mathbf{a}^{(l-k)}& l \geq 1
\end{array}
\right.
\end{equation}
where $\mathbf{w}^{(n)}$ is the eigenvector of $\mathcal{H}_0$ with eigenvalue $\lambda_n$, i.e. $\mathcal{H}_0 \mathbf{w}^{(n)}= \lambda_n \mathbf{w}^{(n)}$. Note that, since for any couple of indices $n$ and $m$ the difference $\lambda_n-\lambda_m$ is not a multiple of $\omega$, i.e. $\tilde{\lambda}_n \neq \tilde{\lambda}_m$, the matrix $( \lambda_n - l \omega - \mathcal{H}_0)$ entering on the right hand side of Eq.(13) is not singular and its inverse matrix is well defined. Therefore, the $N$ quasi-energies of $\mathcal{H}(t)$ are the $N$ eigenvalues $\lambda_n$ of $\mathcal{H}_0$. Since the quasi-energies are defined apart from multiples of $\omega$, we can fold the eigenvalues $\lambda_n$ of $\mathcal{H}_0$ inside the range $(-\omega/2, \omega/2)$, thus yielding $\mu_n=\tilde{\lambda}_n$ for the quasi-energies. 
This shows that the quasi-energy spectrum is entirely real. Note also that, since for any couple of indices $n$ and $m$ the difference $\lambda_n-\lambda_m$ is not a multiple of $\omega$, the quasi-energies are distinct, which excludes the existence of Floquet EPs.\\
\\
{\bf Theorem 2.}  Let $\mathcal{H}(t)$ be a periodic Hamiltonian of the form (9) with period $T= 2 \pi / \omega$ and let us assume that:\\
(i) The eigenvalues $\lambda_n$ of $\mathcal{H}_0$ are real and distinct, with $\lambda_1 < \lambda_2<...< \lambda_N$.\\
(ii) There is a subset of $M$ eigenvalues of $\mathcal{H}_0$, say $\lambda_{y_1}<\lambda_{y_2}<...< \lambda_{y_M}$ with $2 \leq M \leq N$ and $y_{\alpha} \in (1,2,...,N)$ ($\alpha=1,2,...,M$), such that the difference $\lambda_{y_\alpha}-\lambda_{y_\beta}$ is an integer multiple of $\omega$ \footnote{In atomic physics context, such as in laser-driven multilevel atoms, the condition that $\lambda_{y_{\alpha}}-\lambda_{y_{\beta}}$ is an integer multiple of the driving frequency $\omega$ is generally referred to as a multi-photon resonance condition. Here $\lambda_n$ are the energy levels of the atom, whereas $\omega$ is the frequency of the laser field.} ($\alpha,\beta=1,2,...,M$, $ \alpha \neq \beta$), whereas the difference $\lambda_n-\lambda_m$ is not an integer multiple of $\omega$ whenever either one (or both) indices $n$ and $m$ do not belong to the subset $y_1,y_2,...,y_M$.\\
Then the quasi-energy spectrum of $\mathcal{H}(t)$ is entirely real and given by $\mu_n=\tilde{\lambda_n}$, where $\tilde{\lambda}_n$ is the eigenvalue $\lambda_n$ folded inside the range $(-\omega/2, \omega/2)$. The coalescence of the $M$ quasi-energies $\mu_{y_1}=\mu_{y_2}=...\mu_{y_M}$ corresponds to the simultaneous coalescence of their Floquet eigenstates, i.e. $\mathcal{H}(t)$ shows a Floquet EP of order $M$.\\

{\it Proof. } We can proceed like in previous proof of Theorem 1 looking for a solution to Eq.(1) of the Floquet type. Such a solution is of the form (11) with $\lambda=\lambda_n$ and with vectors $\mathbf{a}^{(l)}$ formally given by Eq.(13). If $\lambda_n$ is not any of the eigenvalue $\lambda_{y_{\alpha}}$ of the subset, the matrix $( \lambda_n - l \omega - \mathcal{H}_0)$ entering on the right hand side of Eq.(13) is not singular for any $l$ and its inverse matrix is well defined. Therefore, like in Theorem 1, Eqs.(11) and (13) define a Floquet eigenstate of $\mathcal{H}(t)$ with quasi-energy $\mu_n=\tilde{\lambda}_n$. However, care should be paid when $\lambda_n$ belongs to the subset $\lambda_{y_{\alpha}}$, since in this case the matrix $( \lambda_n - l \omega - \mathcal{H}_0)$ can become singular.\\Let us consider, as a first case, a counter-clockwise loop, corresponding to a negative frequency $\omega<0$. Clearly, for $\lambda_n=\lambda_{y_M}$ (the largest eigenvalue of the subset), 
the matrix $( \lambda_n - l \omega - \mathcal{H}_0)$ is never singular for any $l=1,2,3,...$. Hence $\tilde{\lambda}_{y_M}$ is a quasi-energy and the corresponding Floquet eigenstate is again defined by Eqs.(11) and (13). Conversely, for any other eigenvalue $\lambda_n=\lambda_{y_{\alpha}}$ of the subset ($\alpha=1,2,...,M-1$), 
the matrix $( \lambda_{y_{\alpha}} - l \omega - \mathcal{H}_0)$ becomes singular for some positive integer $l$. In particular, let $G$ be  the positive integer such that $\lambda_{y_M}-\lambda_{y_\alpha}=-G \omega$, so that $( \lambda_{y_{\alpha}} - l \omega - \mathcal{H}_0)$ is not singular for any $l \geq G+1$ while it is singular at $l=G$ (and possibly also at some smaller indices $l$). To avoid the occurrence of the singularity, we should replace in Eq.(13) $\mathbf{w}^{(n)}$ by $ \epsilon \mathbf{w}^{(n)}$ and taking the limit $\epsilon \rightarrow 0$ as the quasi-energy crossing point $\lambda_{y_{M}}-\lambda_{y_\alpha} \simeq -G \omega$ is approached. In this way, the dominant (non-vanishing) terms of $\mathbf{a}^{(l)}$ are those with indices $l \geq G$. Moreover, since for any matrix $\mathcal{M}$ close to singularity the vector $\mathcal{M}^{-1} \mathbf{a}$ is almost parallel to the eigenvector of $\mathcal{M}$ with vanishingly eigenvalue, independently of the choice of $\mathbf{a}$, it readily follows that $\mathbf{a}^{(G)}$ in Eq.(13) becomes parallel to $\mathbf{w}_{\lambda_{y_M}}$. This means that the Floquet eigenstate at the quasi-energy $\mu=\tilde{\lambda}_{y \alpha}$ collapses to the one of the degenerate quasi-energy $\mu=\tilde{\lambda}_{y_M}$. Therefore, $\mathcal{H}(t)$ has a Floquet EP of order $M$, since simultaneous coalescence of $M$ quasi-energies and corresponding Floquet eigenstates is found. Since the quasi-energy $\mu=\tilde{\lambda}_{y_M}$ is defective, one can introduce a set of generalized Floquet eigenstates that restore the completeness of Floquet eigenstates of $\mathcal{H}(t)$. The analytical form of generalized Floquet eigenstates is given in Appendix A.\\
As a second case, let us consider a clockwise loop, corresponding to a positive frequency $\omega>0$.  Clearly, for $\lambda_n=\lambda_{y_1}$ (the smallest eigenvalue of the subset), the matrix $( \lambda_n - l \omega - \mathcal{H}_0)$ is never singular for any $l=1,2,3,...$. Hence $\tilde{\lambda}_{y_1}$ is a quasi-energy and the corresponding Floquet eigenstate is again defined by Eqs.(11) and (13). On the other hand, for any other eigenvalue $\lambda_n=\lambda_{y_{\alpha}}$ of the subset ($\alpha=2,3,...,M$), the matrix $( \lambda_{y_{\alpha}} - l \omega - \mathcal{H}_0)$ becomes singular for some positive integer $l$. Following the same reasoning as in the previous case of  a counter-clockwise loop, the Floquet eigenstates corresponding to such quasi-energies collapse to the Floquet eigenstate defined by Eqs.(11) and (13) with $\lambda_n=\lambda_{y1}$. Hence, the degeneracy of the quasi-energies $\tilde{\lambda}_{y_{\alpha}}$ corresponds again to the simultaneous coalescence of the corresponding Floquet eigenstates, i.e. there is a Floquet EP of order $M$. However, by reversing the sign of $\omega$, i.e. the circulation direction of the loop $\mathcal{C}$, the Floquet eigenstates collapse to a different state: the state emanating from the highest eigenvalue $\lambda_{y_M}$ for a counter-clockwise loop ($\omega<0$), the state emanating from the lowest eigenvalue $\lambda_{y1}$ for a clockwise loop ($\omega>0$). As we will discuss in following Sec.3.4, the different collapse of Floquet eigenstates in the two circulation directions of the loop is responsible for the chirality of the system, i.e. the selection of different final states, starting from some common initial state, when the loop $\mathcal{C}$ is slowly and repeatedly circulated in the two opposite directions. A schematic diagram that illustrates the results of Theorem 2 is shown in Fig.1. 
\begin{figure}[htb]
\centerline{\includegraphics[width=13cm]{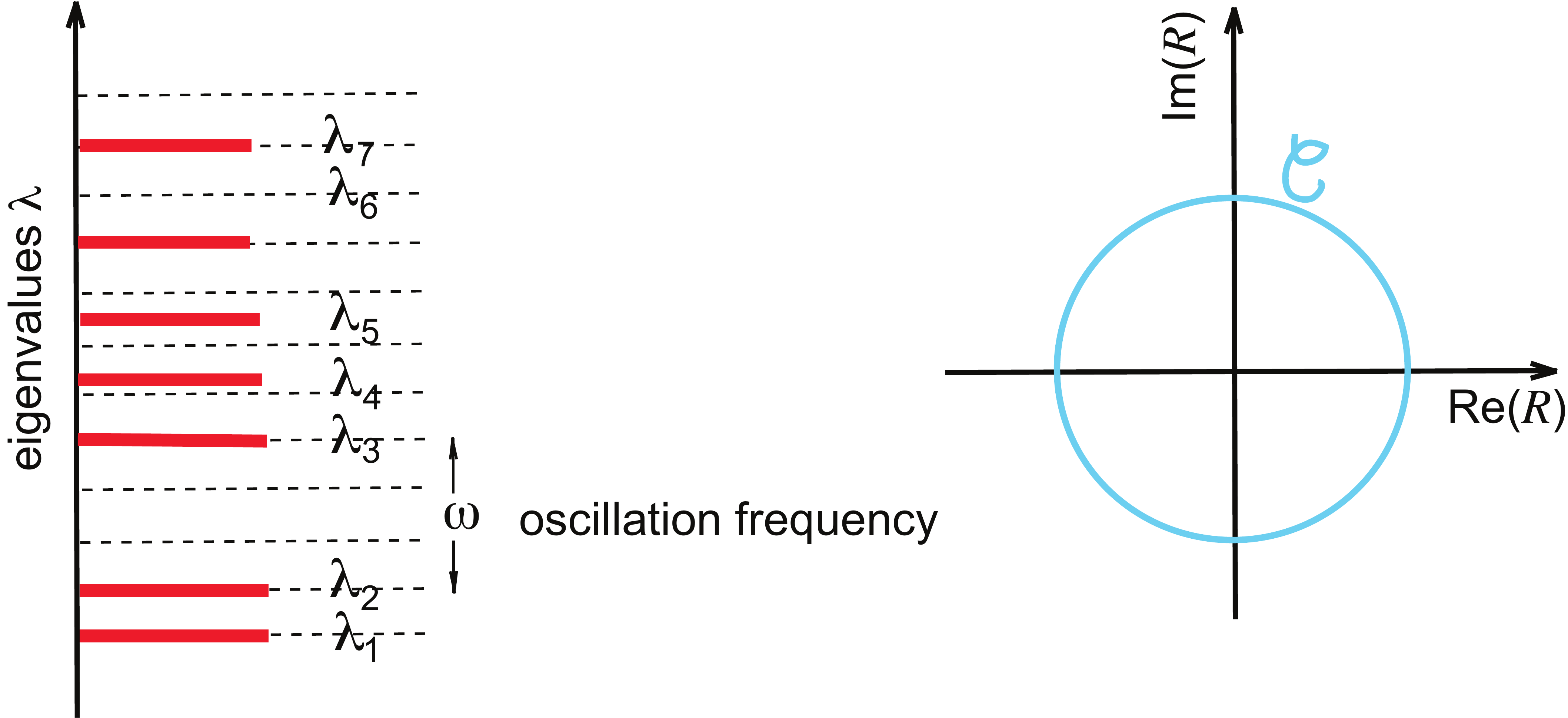}} \caption{Schematic diagram showing the appearance of a Floquet EP under the conditions stated in Theorem 2. The Hamiltonian $\mathcal{H}_0$ has $N$ distinct and real eigenvalues $\lambda_1$, $\lambda_2$, ..., $\lambda_N$, with $N=7$ in the case illustrated in the figure. Given the spacing of eigenvalues shown in the left panel of the figure and the value of frequency $\omega$ at which the complex parameter $R(t)$ oscillates, describing either a clockwise or counter-clockwise closed loop $\mathcal{C}$ (right panel), for the subset of $M=3$ eigenvalues $\lambda_{y_{\alpha}}$, with $y_{1}=2$, $y_{2}=3$ and $y_{3}=7$, the difference $\lambda_{y_{\alpha}}-\lambda_{y_{\beta}}$ is a multiple of $\omega$ ($\lambda_3-\lambda_2= \omega$, $\lambda_7-\lambda_2=3 \omega$, $\lambda_7-\lambda_3= 2 \omega$). According to Theorem 2, a Floquet EP of order $M=3$ is found. Three among the seven Floquet eigenstates of $\mathcal{H}(t)$ collapse to the same state, namely the state emanating from the largest eigenvalue $\lambda_7$ of the subset for a counter-clockwise loop ($\omega < 0$), the state emanating from the lowest eigenvalue  $\lambda_2$ of the subset for a clockwise loop ($\omega > 0$).}
\end{figure}

\subsection{Adiabatic analysis}
We  are especially interested to consider the slow limit $\omega \rightarrow 0$ of cycling, where the dynamics over one oscillation cycle is well approximated by an adiabatic analysis. In particular, we wish to establish a correspondence between the exact Floquet eigenstates of $\mathcal{H}(t)$, as defined in Theorems 1 and 2 above, and the
quasi-periodic adiabatic states of $\mathcal{H}(t)$.  To this aim, let us indicate by $\sigma_n(t)$ and $\mathbf{e}^{(n)}(t)$ the instantaneous eigenvalues and corresponding eigenvectors of $\mathcal{H}(t)$, i.e.
\begin{equation}
\mathcal{H}(t) \mathbf{e}^{(n)}(t)
= \sigma_n(t) \mathbf{e}^{(n)}(t).
\end{equation}
We assume that the instantaneous eigenvalues $\sigma_n(t)$ are distinct and separated each other by a finite gap over the entire oscillation cycle. Therefore, at each time $\mathbf{e}^{(n)}(t)$ form a complete basis. The eigenvalues $\sigma_n(t)$ are uniquely determined at each time by the determinatal equation, which is an algebraic equation of order $N$ with complex coefficients that vary periodically in time with period $T$. Rather generally $\sigma_n(t)$ and corresponding eigenvector $\mathbf{e}^{(n)}(t)$ are  not single-valued functions of time over one oscillation cycle, i.e. if we continuously follow the change of the eigenvalue $\sigma_n(t)$, as the vector  $\mathbf{R}(t)$ continuously change from $t=0$ to $t=T$ describing the loop $\mathcal{C}$ either clockwise or counter-clockwise, we have rather generally a flip of eigenvalues and corresponding eigenvectors at the end of the cycle, i.e. $\sigma_n(T^-) = \sigma_m(0^+)$ for some $m \neq n$. Only after $N$ successive cycles the  condition $\sigma_n(NT^-)=\sigma_n(0^+)$ is met. This is because the Fourier series of $\sigma_n(t)$ and $\mathbf{e}^{(n)}(t)$ generally show sub-harmonic terms $\sim \exp(i \omega t/N)$.  For example, for a two-state system ($N=2$) the flip of eigenvalues and corresponding eigenvectors is known to occur whenever the path in parameter space, described by a complex parameter  $R=R(t)$, encircles an EP, i.e. a point $R=R_0$ at which the stationary Hamiltonian $\mathcal{H}$ shows a static EP \cite{r16,r17,r18,r25,r36,r37,r38,r40}. Here we avoid such a case and assume that each eigenvalue $\sigma_n(t)$ and corresponding eigenvector $\mathbf{e}^{(n)}(t)$ are single-valued functions of time over each cycle, i.e. $\sigma_n(T^-)=\sigma_n(0^+)$ and $\mathbf{e}^{(n)}(T^-)=\mathbf{e}^{(n)}(0^+)$. Since the elements of $\mathcal{H}(t)$ are assumed to be periodic in time and composed by positive (or negative) frequency components solely, from the determinantal equation of the eigenvalues it readily follows that $\sigma_n(t)$ is composed by a mean value plus positive (or negative) frequency components solely, i.e. in the Fourier series of $\sigma_n(t)$ all negative (or positive) frequency terms vanish. This implies that the time average of the instantaneous eigenvalue over one oscillation cycle should coincide with one eigenvalue of the stationary matrix $\mathcal{H}_0$, i.e.
\begin{equation}
\frac{1}{T}\int_0^T dt \sigma_n(t)=\lambda_n.
\end{equation}
To perform an adiabatic analysis of Eq.(1) in the slow-cycling limit, let us expand the state vector $\mathbf{a}(t)$ in series of the instantaneous eigenvectors $\mathbf{e}^{(n)}(t)$ of $\mathcal{H}(t)$, i.e.
\begin{equation}
\mathbf{a}(t)=\sum_{n=1}^N f_n(t) \mathbf{e}^{(n)}(t) \exp \left[ -i \int_0^t d\xi \sigma_n(\xi) \right]
\end{equation}
with time-dependent amplitudes $f_n(t)$. Indicating by $\mathbf{e}^{(n) \dag}(t)$ the instantaneous eigenvector of the adjoint matrix $\mathcal{H}^{\dag}(t)$ with eigenvalue $\sigma_n^*(t)$, i.e. $\mathcal{H}^{\dag}(t) \mathbf{e}^{(n) \dag} (t) =\sigma^*_n(t) \mathbf{e}^{(n) \dag} (t)$, taking into account that the scalar product $\langle \mathbf{e}^{(n) \dag} (t) | \mathbf{e}^{(m)}(t) \rangle$ vanishes for $n \neq m$, the evolution equations for the amplitudes $f_n(t)$ are readily found after substitution of the Ansatz (16) into Eq.(1) and taking the scalar product by $\langle \mathbf{e}^{(n) \dag}(t)|$. One obtains
\begin{eqnarray}
\frac{df_n}{dt} & =& -\frac{1}{\langle \mathbf{e}^{(n) \dag}(t) | \mathbf{e}^{(n)}(t) \rangle} \sum_{m=1}^{N} \langle \mathbf{e}^{(n) \dag} (t) | \frac{d \mathbf{e}^{(m)}}{dt} \rangle f_m(t) \nonumber \\
& \times & \exp \left[i  \int_0^t d \xi (\sigma_n(\xi)-\sigma_m(\xi) )\right]
\end{eqnarray}
which are exact equations. In the slow cycling limit $\omega \rightarrow 0$, the adiabatic approximation is obtained by neglecting the coupling terms of amplitudes in Eq.(17), i.e. by assuming
\begin{equation}
\frac{df_n}{dt} \simeq -\frac{\langle \mathbf{e}^{(n) \dag} (t) | \frac{d \mathbf{e}^{(n)} }{dt} \rangle}{\langle \mathbf{e}^{(n) \dag}(t) | \mathbf{e}^{(n)} (t) \rangle} f_n
\end{equation}
which yields
\begin{equation}
f_n(t) \simeq f_n(0) \exp \left[ -\int_0^t  d \xi \frac{\langle \mathbf{e}^{(n) \dag}(\xi) | \frac{d \mathbf{e}^{(n)}}{d \xi} \rangle}{\langle \mathbf{e}^{(n) \dag}(\xi) | \mathbf{e}^{(n)}(\xi) \rangle}  \right] \equiv f_n(0) \exp [-i \varphi_{n}(t)] . 
\end{equation}
Note that the change of phase and amplitude of $f_n$ in the adiabatic approximation arises from the usual complex Berry phase term $\varphi_n(t)$ \cite{r17,r45,r46,r46bis,r47}. After one period, i.e. after the parameter vector $\mathbf{R}(t)$ has described a closed loop $\mathcal{C}$ in complex parameter space, the contribution of the complex Berry phase takes the form
\begin{equation}
 \varphi_n (T)=-i 
\int_0^T d \xi \frac{\langle \mathbf{e}^{(n) \dag} (\xi) | \frac{d \mathbf{e}^{(n)}}{d \xi} \rangle}{\langle \mathbf{e}^{(n) \dag}(\xi) | \mathbf{e}^{(n)}(\xi) \rangle} =
-i  \oint_{\mathcal{C}} \frac{\langle \mathbf{e}^{(n) \dag} (\mathbf{R}) | d \mathbf{e}^{(n)}(\mathbf{R}) \rangle}{\langle \mathbf{e}^{(n) \dag}(\mathbf{R}) | \mathbf{e}^{(n)}(\mathbf{R}) \rangle}.
\end{equation}
Under the conditions discussed above that $\lambda (\mathbf{R})$ and $\mathbf{e}_n(\mathbf{R})$ are not multi-valued functions of $\mathbf{R}$ and that the loop $\mathcal{C}$ does not encircle an EP, the integral over the closed path $\mathcal{C}$ vanishes, i.e. $\varphi_n(T)=0$. For example, if we assume a single complex parameter $R(t)$ [$k=1$ in Eq.(9)], the integral on the right hand side of Eq.(20) can be computed as a contour integral in complex $R$ plane. Since the path $\mathcal{C}$ does not encircle an EP, the function under the sign of integral in Eq.(20) in holomorphic in the domain internal to the loop $\mathcal{C}$ because the denominator  ${\langle \mathbf{e}^{(n) \dag}(\mathbf{R}) | \mathbf{e}^{(n)}(\mathbf{R}) \rangle}$ does not vanish (it vanishes only at an EP). Therefore, owing to the Cauchy integral theorem, the integral (20) along the closed path $\mathcal{C}$ vanishes, i.e. $\varphi_n(T)=0$.\\ 
We mention that, for slow cycling ($\omega$ much smaller than the gap separating the instantaneous eigenvalues $\sigma_n(t)$), the adiabatic approximation obtained by neglecting the rapidly-oscillating terms in Eq.(17) (rotating-wave approximation) is accurate to describe the dynamical evolution of there system up to a time of order $\sim 1 / \omega$, i.e. over one (or a few) oscillation cycles. This can be formally proven by a multiple time scale asymptotic analysis of Eq.(17), where the rotating-wave approximation is shown to describe the evolution of amplitudes $f_n$ in the lowest-order time scale $ \sim 1 /\omega$. However, the rotating-wave approximation can not be applied to describe the evolution of amplitudes over time scales longer than $ \sim 1 / \omega$, i.e. breakdown of the adiabatic theorem can arise when repeated circulation of the loop $\mathcal{C}$ is considered owing to the appearance of resonances in the asymptotic analysis. Let us first consider the system dynamics over one oscillation cycle, i.e for one circulation of the loop $\mathcal{C}$, so as the adiabatic approximation can be applied. In this case, note that the set of $N$ adiabatic functions 
\begin{equation}
\mathbf{p}^{(n)}(t)= \mathbf{e}^{(n)}(t) \exp \left[ -i \varphi_n(t)-i \int_0^t d \xi \sigma_n( \xi) \right]
\end{equation}
satisfy the conditions
\begin{equation}
\mathbf{p}^{(n)}(t+T)=\mathbf{p}^{(n)}(t) \exp(-i \mu_n T)
\end{equation}
with
\begin{equation}
\mu_n T= \left[ \varphi_n(T)+\int_0^T dt \sigma_n(t) \right]= \lambda_n T 
\end{equation}
where we used Eq.(15) and $\varphi_n(T)=0$. Equations (22) and (23) clearly show that, in the adiabatic limit, the functions $\mathbf{p}^{(n)}(t)$ defined by the instantaneous eigenvectors of $\mathcal{H}(t)$ [Eq.(21)] are the Floquet eigenvectors of $\mathcal{H}(t)$ with quasi-energies $\mu_n=\tilde{\lambda}_n$. We therefore retrieve the (exact) result, shown in Theorems 1 and 2, that the quasi-energy spectrum of $\mathcal{H}(t)$ is entirely real and given by the eigenvalues of the stationary Hamiltonian $\mathcal{H}_0$, folded into the interval $(-\omega/2, \omega/2)$. If the conditions stated in Theorem 1 are satisfied, i.e. if the quasi-energies are distinct and there is not a Floquet EP, clearly $\mathbf{p}^{(n)}(t)$ should provide an approximate form of the distinct and linearly-independent exact Floquet eigenvectors $\mathbf{f}^{(n)}(t)$ of $\mathcal{H}(t)$, defined by the Fourier series Eqs.(11) and (13), i.e one has $\mathbf{f}^{(n)}(t) \simeq \mathbf{p}^{(n)}(t)$ for any $n=1,2,...,N$. However, some care should be paid when the conditions stated in Theorem 2 are met, i.e. in the presence of quasi-energy degeneracy leading to the appearance of a Floquet EP of order $M$. In this case, the set of adiabatic states $\{ \mathbf{p}^{(n)}(t) \}$ are always linearly-independent and of Floquet form, i.e. each of them reproduces itself, apart from the phase term $\sim \exp(-i \mu_n T)$, after one oscillation cycle. On the other hand, at a Floquet EP the exact Floquet states $\mathbf{f}^{(y_1)}(t)$, $\mathbf{f}^{(y_2)}(t)$, ..., $\mathbf{f}^{(y_M)}(t)$ coalesce and to restore the completeness of eigenstates we must introduce the generalized Floquet eigenstates $\mathbf{F}^{(y_n)}(t)$, which are not of the Floquet form for $n=2,3,..,M$ since they contain secular growing terms in time [see Eq.(6)]. The prediction of the adiabatic analysis is thus seemingly at odd with the exact Floquet theory in case of a Floquet EP.  Nevertheless, the contradiction is removed after observing that, in the slow-cycling limit $\omega \rightarrow 0$, the ratio between the elements of vectors $\mathbf{Q}^{(y_n)}$ and $\mathbf{q}^{(y_1)}$ entering in Eq.(6) is large and of order $\sim 1/ \omega^{(2n-2)}$, with $n=2,3,..,M$ \footnote{Such  a property can be readily proven as follows. For the sake of definiteness, let us consider the worst case where all quasi energies collapse, i.e. $M=N$ corresponding to a Floquet EP of order $N$. Let us indicate by $\mathcal{G}$ the $N \times N$ matrix whose column vectors are the instantaneous eigenvectors $\mathbf{e}^{(1)}(0)$, $\mathbf{e}^{(2)}(0)$,..., $\mathbf{e}^{(N)}(0)$  of $\mathcal{H}$ at initial time $t=0$.  According to the adiabatic analysis, in the slow cycle limit the solution to Eq.(1) with the initial condition $\mathbf{a}(0)=\mathbf{e}^{(n)}(0)$ evolves approximately into the state $\mathbf{e}^{(n)}(0) \exp(-i \lambda_n T)=\mathbf{e}^{(n)}(0) \exp(-i \tilde{\lambda}_n T)$ after one oscillation cycle, with an error of order $\sim \omega$ as compared to the exact solution. Hence, from Eq.(2) with $t=T$ one obtains $\exp(-i \mathcal{R}T) \mathcal{G}=\mathcal{G} \exp(-i \tilde{\Lambda} T)+O(\omega)$, where $\tilde{\Lambda}$ is the diagonal matrix of the folded eigenvalues $\tilde{\lambda}_n$, i.e. of quasi-energies $\mu_n=\tilde{\lambda}_n$, and the error is of order $\sim \omega$. The matrix $\mathcal{R}$ can be thus written as $\mathcal{R}= \mathcal{G}  \tilde{ \Lambda } \mathcal{G}^{-1} +O(\omega^2)$, where $O(\omega^2)$ indicates a matrix whose elements, as compared to the fist term, are small and of order $\sim \omega^2$. Since the quasi-energies collapse to the same value $\mu_n=\mu$, one has $\tilde{\Lambda}= \mu \mathcal{I}$ ($\mathcal{I}$ is the identity matrix).
Therefore, one can write $\mathcal{R}=\mathcal{G}  \mathcal{J} \mathcal{G}^{-1}$, where $\mathcal{J}=\mu \mathcal{I}+\omega^2 \mathcal{M}$ and the elements of the matrix $\omega^2 \mathcal{M}$ are small of order $\sim \omega^2$. The set of $N$ generalized eigenvectors $\mathbf{Q}_{n}$ of $\mathcal{R}$ are obtained from those $\mathbf{z}_{n}$ of $\mathcal{J}$ by application of the transformation $\mathcal{G}$, i.e.  $ \mathbf{Q}_{n} = \mathcal{G} \mathbf{z}_{n}$.  The eigenvectors  $\mathbf{z}_{n}$ are the solutions of the cascaded equations
$(\mathcal{J}-\mu) \mathbf{z}_{1}=0$, $(\mathcal{J}-\mu) \mathbf{z}_{2}=\mathbf{z}_{1}$,...,  $(\mathcal{J}-\mu) \mathbf{z}_{N}=\mathbf{z}_{N-1}$, i.e. 
$ \mathcal{M} \mathbf{z}_{1}=0$, $\mathcal{M} \mathbf{z}_{2}=\mathbf{z}_{1} / \omega^2$,...,  $\mathcal{M} \mathbf{z}_{N}=\mathbf{z}_{N-1} / \omega^2$. Therefore, when solving the equation $\mathcal{M} \mathbf{z}_{n}= \mathbf{z}_{n-1} / \omega^2$, it readily follows that the ratio between the elements of $\mathbf{z}_{n}$ and $\mathbf{z}_{n-1}$ is large and of order $ \sim 1/ \omega^2$. The same scaling holds for the generalized eigenvectors $\mathbf{Q}_n= \mathcal{G} \mathbf{z}_n$ of $\mathcal{R}$.} 
 Therefore, over a single oscillation cycle, i.e. after the time $t=T$, one has $\mathbf{F}^{(y_n)}(t+T) \simeq \mathbf{F}^{(y_n)}(t) \exp(-i \mu_{y_n}T)$ also for  the generalized Floquet eigenstates, the secularly growing terms entering in Eq.(6) becoming dominant only after several oscillation cycles and being negligible in the time interval $(0,T)$. Hence the adiabatic states (21) provide an approximate form of the entire set of Floquet eigenstates (including the generalized ones) even at an EP. 
 \subsection{Multi-cycling chirality induced by Floquet EPs}
Since the set of adiabatic states (21) provide an accurate approximation of the entire set of Floquet eigenvectors of $\mathcal{H}(t)$ over one oscillation cycle, when the system dynamics is observed over one (or a few) oscillation cycles nonadiabatic transitions are not observed. However, in presence of a Floquet EP the adiabatic states cannot describe the secular growing terms arising in the generalized Floquet eigenstates $\mathbf{F}^{(y_n)}(t)$ [Eq.(6) with $n=2,3,...,M$]. While such terms remain small and can be therefore neglected when the system is cycled once or a few times, they cannot be neglected when the system is cycled many times around the loop $\mathcal{C}$. In other words, nonadiabatic (coupling) terms in Eq.(17) can not be neglected when the dynamics is observed at time scales longer than $\sim T$, and mixing of instantaneous adiabatic states should occur in the presence of a Floquet EP. Such a breakdown of the adiabatic  theorem basically stems from the appearance of resonances in a multiple time scale asymptotic analysis of Eq.(17).
Interestingly, for a counter-clockwise loop, i.e. for $\omega<0$, the Floquet eigenstate $\mathbf{f}^{(y_M)}(t)$ of $\mathcal{H}(t)$, which is the dominant state of the dynamics, corresponds to the adiabatic state $\mathbf{p}^{(y_M)}(t)$, whereas for  a clockwise loop, i.e. for $\omega>0$, one has $\mathbf{f}^{(y_1)}(t) \simeq \mathbf{p}^{(y_1)}(t)$. This property readily follows from the arguments used in the proofs of Theorems 1 and 2. The different selection of dominant Floquet state when the Hamiltonian is cycled clockwise or counterclockwise introduces a chiral behavior in the system, i.e. the selection of a distinct final state, starting from the same initial state, depending on the circulation direction of the loop. For example, let us prepare the system at $t=0$ in the instantaneous eigenvector $\mathbf{e}^{(y_1)}(0)$ of $\mathcal{H}(0)$. If the system is cycled one (or few) times either clockwise or counter-clockwise, the system basically returns to its initial state, i.e. nonadiabatic transitions are prevented and there is not any chirality in the dynamics. However, after many oscillation cycles, a chiral behavior emerges: while for a clockwise cycling the system remains in the same state and no adiabatic transitions occur (this is because $\mathbf{e}^{(y_1)}(0)$ is the dominant Floquet eigenstate), for counter-clockwise cycling the system is dominated by the other state  $\mathbf{e}^{(y_M)}(0)$ after several oscillation cycles, indicating breakdown of adiabatic following (this is because $\mathbf{e}^{(y_M)}(0)$ is now the dominant state of the dynamics). An example of the chiral behavior induced by a Floquet EP is discussed in the next section.
 
\section{Floquet exceptional points and chirality: an example}
As an illustrative example of chirality induced by a Floquet EP, let us consider the $3 \times 3$ Hamiltonian
\begin{equation}
\mathcal{H}(t)= \mathcal{H}_0+R(t) \mathcal{H}_1
\end{equation}
with 
\begin{equation}
\mathcal{H}_0 = \left(
\begin{array}{ccc}
0 & 1 & 0 \\
\Omega^2/2 & 0 & 1 \\
0 & \Omega^2/2 & 0
\end{array}
\right) \; ,\;\; 
\mathcal{H}_1 = \left(
\begin{array}{ccc}
0 & 0 & 0 \\
1 & 0 & 0 \\
0 & -1 & 0
\end{array}
\right)
\end{equation}
where $\Omega$ is a real constant parameter. The Hamiltonian periodically varies in time via the complex parameter 
\begin{equation}
R(t)=R_0 \exp(i \omega t)
\end{equation}
which describes a closed loop $\mathcal{C}$ in complex $R$ plane after a time interval $T$. The eigenvalues $\lambda_n$ of $\mathcal{H}_0$ are given by
 \begin{equation}
 \lambda_1=-\Omega, \;\; \lambda_2=0, \;\; \lambda_3=\Omega
 \end{equation}
Interestingly, the instantaneous eigenvalues $\sigma_n(t)$ of $\mathcal{H}(t)$ are independent of parameter $R$, i.e. of time, and are the same as those of $\mathcal{H}_0$, i.e. 
 \begin{equation}
 \sigma_1(t)= -\Omega \; \; \sigma_2(t)=0, \;\; \sigma_3(t)=\Omega
  \end{equation}
 whereas the corresponding eigenvectors are given by
 
\begin{figure}[htb]
\centerline{\includegraphics[width=15cm]{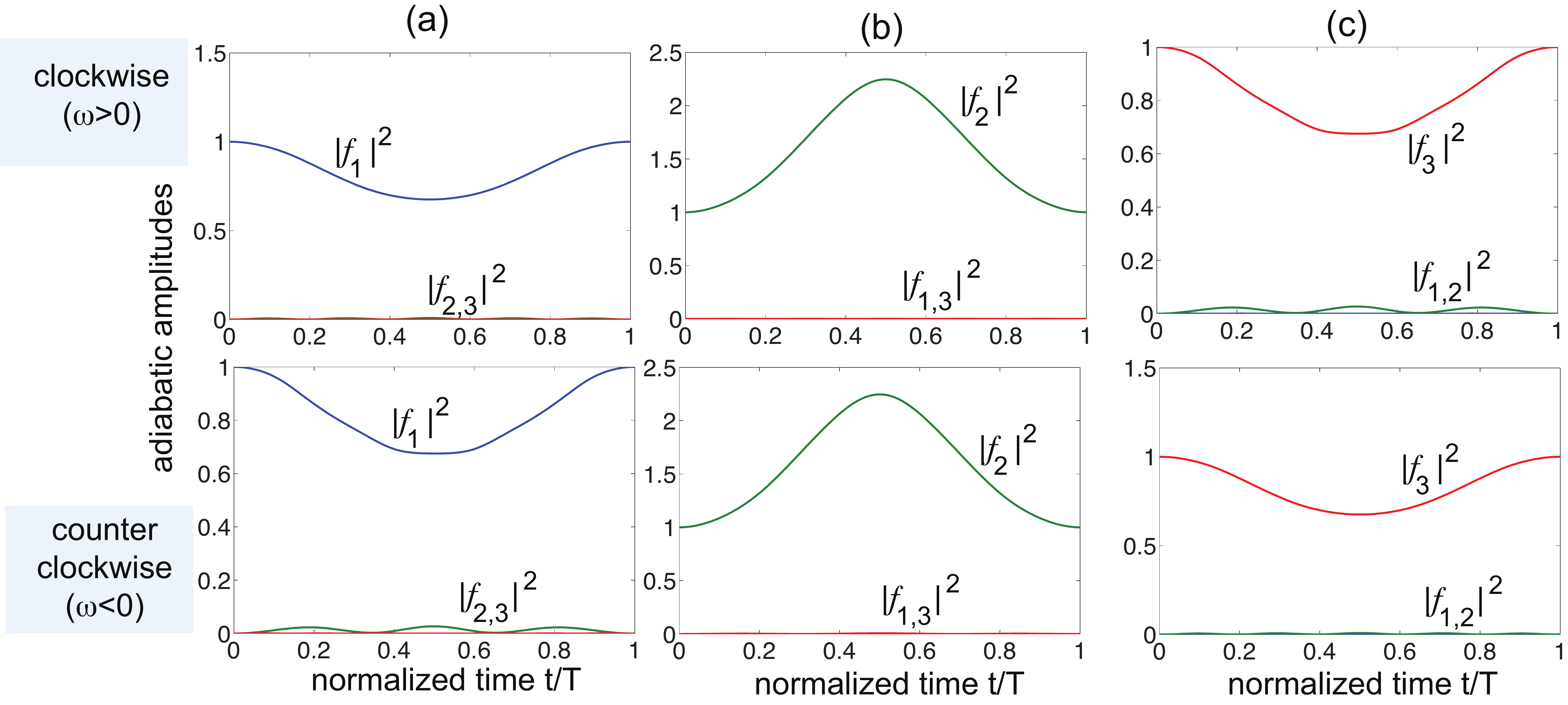}} \caption{
Evolution of the adiabatic amplitudes $|f_n(t)|^2$ ($n=1,2,3$) as obtained by numerical simulations of Eqs.(31-33) using an accurate variable-step fourth-order Runge-Kutta method for parameter values $\Omega=1$, $R_0=0.2$ and $| \omega|=0.25$ (corresponding to the even multiphoton resonance condition $2 \Omega/  \omega=8$). The initial condition is $f_1(0)=1, f_{2,3}(0)=0$ in (a), $f_2(0)=1, f_{1,3}(0)=0$ in (b)  and $f_3(0)=1, f_{1,2}(0)=0$ in (c). The upper panels refer to a clockwise cycle ($\omega>0$), whereas the lower panels to a counter-clockwise loop ($\omega<0$).}
\end{figure}

\begin{figure}[htb]
\centerline{\includegraphics[width=15cm]{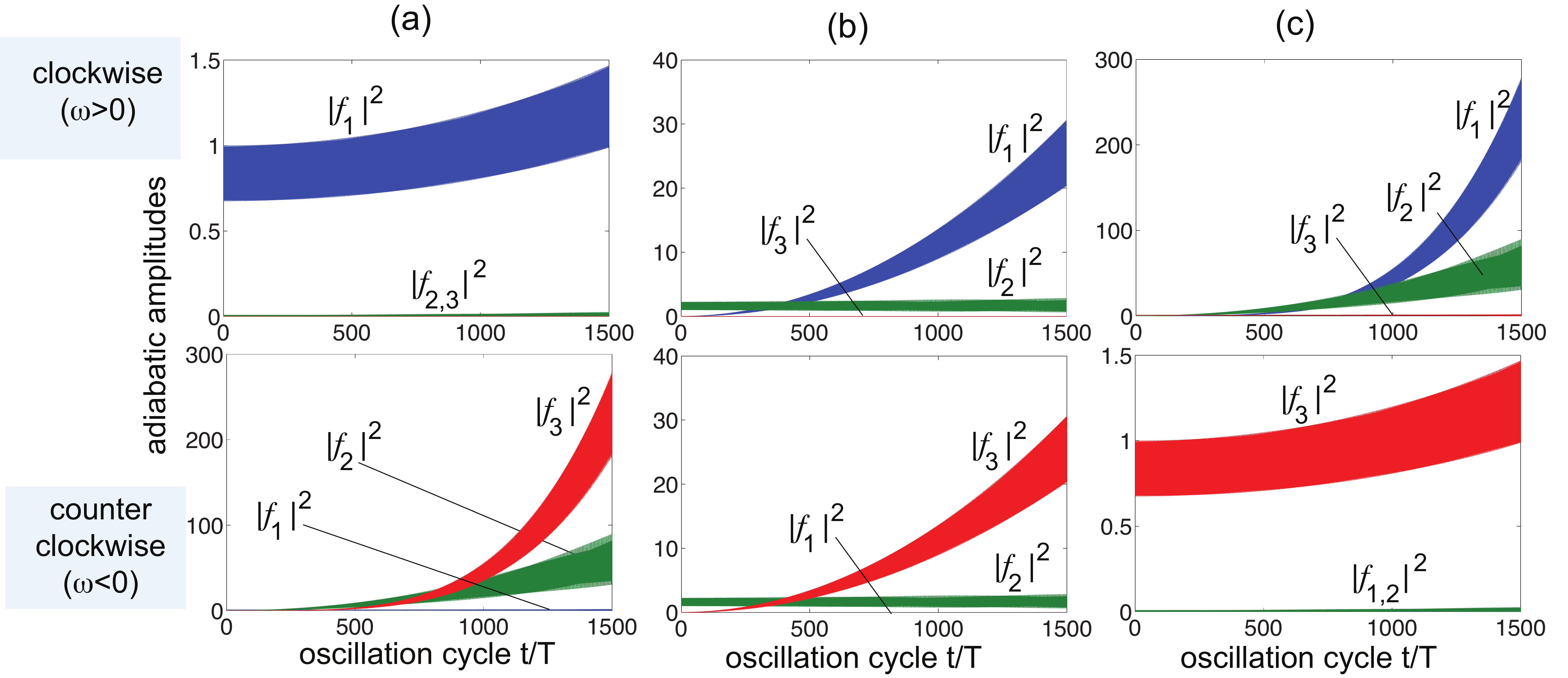}} \caption{
Same as Fig.2, but when the system undergoes repeated cycles.}
\end{figure}

\begin{equation}
\mathbf{e}^{(1)}=
\left(
\begin{array}{c}
1 \\
-\Omega \\
\Omega^2 / 2 -R
\end{array}
\right)  \; , \; 
\mathbf{e}^{(2)}=
\left(
\begin{array}{c}
1 \\
0 \\
-\Omega^2 / 2 -R
\end{array}
\right) , \; 
\mathbf{e}^{(3)}=
\left(
\begin{array}{c}
1 \\
\Omega \\
\Omega^2 / 2 -R
\end{array}
\right). \;\;\;
\end{equation} 
Note that, since the eigenvalues of $\mathcal{H}$ are independent of the parameter $R$ and are distinct, no static EPs are encircled by $R$ as it describes an arbitrary closed loop in complex plane. Nevertheless, according to the Theorem 2 Floquet EPs can be found for special driving frequencies. In particular:\\
(i) For a driving frequency $\omega$ satisfying the odd resonance condition $2\Omega=n \omega$ with $n$ odd, the two quasi energies $\lambda_1$ and $\lambda_3$ coalesce, corresponding to a second-order Floquet EP.\\
(ii) For a driving frequency $\omega$ satisfying the even resonance condition $2\Omega=n \omega$ with $n$ even, the three quasi energies $\lambda_1$, $\lambda_2$ and $\lambda_3$ coalesce, corresponding to a third-order Floquet EP.\\ 
To illustrate the impact of a Floquet EP on the system dynamics, let us consider the evolution of amplitudes $f_n$ in the basis of instantaneous eigenstates of $\mathcal{H}(t)$ [Eq.(17)]. Taking into account that the adjoint eigenvectors are given by
\begin{equation}
\mathbf{e}^{(1) \dag}=
\left(
\begin{array}{c}
\Omega^2/2+R^* \\
-\Omega \\
1
\end{array}
\right)  \; , \; 
\mathbf{e}^{(2) \dag}=
\left(
\begin{array}{c}
-\Omega^2/2+R^* \\
0 \\
1
\end{array}
\right) , \; 
\mathbf{e}^{(3) \dag}=
\left(
\begin{array}{c}
\Omega^2/2+R^* \\
\Omega \\
1
\end{array}
\right) \;\;\;
\end{equation} 
Eq.(17) takes the form
\begin{eqnarray}
\frac{df_1}{dt} & = & \frac{\dot{R}}{2\Omega^2} \left[ f_1+f_2 \exp(-i \Omega t)+f_3 \exp(-2i \Omega t) \right] \\
\frac{df_2}{dt} & = & -\frac{\dot{R}}{\Omega^2} \left[f_1 \exp(i \Omega t) + f_2+f_3 \exp(-i \Omega t)\right] \\
\frac{df_3}{dt} & = & \frac{\dot{R}}{2\Omega^2} \left[f_1 \exp(2i \Omega t) + f_2 \exp(i \Omega t) +f_3 \right] 
\end{eqnarray}
where $\dot{R} \equiv dR/dt=i \omega R_0 \exp(i \omega t)$. In the slow cycling limit $\omega \rightarrow 0$, the adiabatic approximation basically corresponds to disregard the rapidly-oscillating terms $\exp ( \pm i \Omega t)$ and $\exp( \pm i 2 \Omega t)$ in Eqs.(31-33) that couple the amplitudes of different adiabatic states (rotating-wave approximation). 
This yields
\begin{equation}
f_n(t) \simeq f_n(0) \exp[- i \varphi_n(t)]
\end{equation}
($n=1,2,3$) with complex Berry phases
\begin{equation}
\varphi_1(t)=\varphi_3(t)=-\frac{\varphi_2(t)}{2}=i \frac{R(t)-R(0)}{2 \Omega^2}.
\end{equation}
Note that, according to the theoretical analysis, the Berry phases vanish after each oscillation cycle.
The rotating-wave approximation, being the leading-order approximation in a multiple time scale asymptotic expansion, correctly and safely describes the system dynamics for a time scale up to $\sim 1 / \omega$, i.e. a few oscillation cycles. Far from a multiphoton resonance, i.e. in the absence of a Floquet EP, the rotating-wave approximation generally turns out to correctly describe the system dynamics even at time scales longer than $ \sim 1 / \omega$, indicating that the adiabatic regime is maintained even after several cycles. This is because of the absence of secular terms in the asymptotic analysis arising from resonances. However, if a multiphoton resonance condition is met, i.e. at a Floquet EP, the rotating-wave approximation breaks down at a time scale longer than $ \sim 1 / \omega$, so that nonadiabatic effects, i.e. mixing of amplitudes $f_n$, is observed after several oscillation cycles. Nonadiabatic effects are precisely responsible for the appearance of a chirality in the system dynamics.  Figures 2-5 show a few different dynamical behaviors that are observed by varying the circulation direction of the cycle, the  initial state and the oscillation frequency. Figures 2 and 3 show the typical behavior of the adiabatic amplitudes $|f_n(t)|^2$ observed when the oscillation frequency $\omega$ satisfies an even multiphoton resonance condition, corresponding to the appearance of a third-order EP. In Fig.2 the system undergoes a single cycle, either clockwise (upper panels) or counter-clockwise (lower panels), with the initial condition corresponding to the excitation of one of the three instantaneous eigenstates. Note that, regardless of the circulation direction of the loop and the initial state preparation, after completing the cycle the system returns to its initial state and excitation of other adiabatic amplitudes is negligible. This shows that, over one oscillation cycle, nonadiabatic transitions are negligible and there is not any chiral behavior. Note that the modulus square of the amplitude of the initially excited state is not constant in time, rather it varies over the oscillation cycle because of the complex nature of the Berry phases $\varphi_n(t)$ as a signature of non-Hermitian dynamics. However, since the Berry phase over one oscillation cycle vanishes, at the end of the cycle the system returns to its initial state, i.e. the dynamics is (almost) periodic. The same behavior holds whenever the system undergoes a few oscillation cycles. However the dynamics dramatically changes when the system undergoes several (e.g. a few hundreds) oscillation cycles (Fig.3). Here nonadiabatic transitions are clearly observed and the final dominant state depends on the circulation direction of the loop, i.e. a chiral behavior is found. In particular, according to the theoretical analysis for a clockwise loop ($\omega>0$) the dominant state is $\mathbf{e}^{(1)}$, whereas for a counter-clockwise loop ($\omega<0$) the dominant state is $\mathbf{e}^{(3)}$. A similar behavior is found when the oscillation frequency is set to satisfy an odd multiphoton resonance, corresponding to a second-order EP, as shown in Fig.4. As compared to the case of Fig.3, note that in the latter case chirality is not observed when the system is initially prepared in the state $\mathbf{e}^{(2)}(0)$ [compare Fig.3(b) and 4(b)]. We remark that the chiral behavior is strictly related to breakdown of adiabatic following in one (or both) of the circulation directions induced by the Floquet EP. If the multiphoton resonance condition leading to the appearance of the Floquet EP is not satisfied, the chiral behavior of the dynamics disappears. This is shown, as an example, in Fig.5, where the frequency $\omega$ is slightly changed from $\omega=0.25$, corresponding to the even multiphoton resonance condition $2 \Omega / | \omega|=8$, to $\omega=0.26$.\\ 

\begin{figure}[htb]
\centerline{\includegraphics[width=15cm]{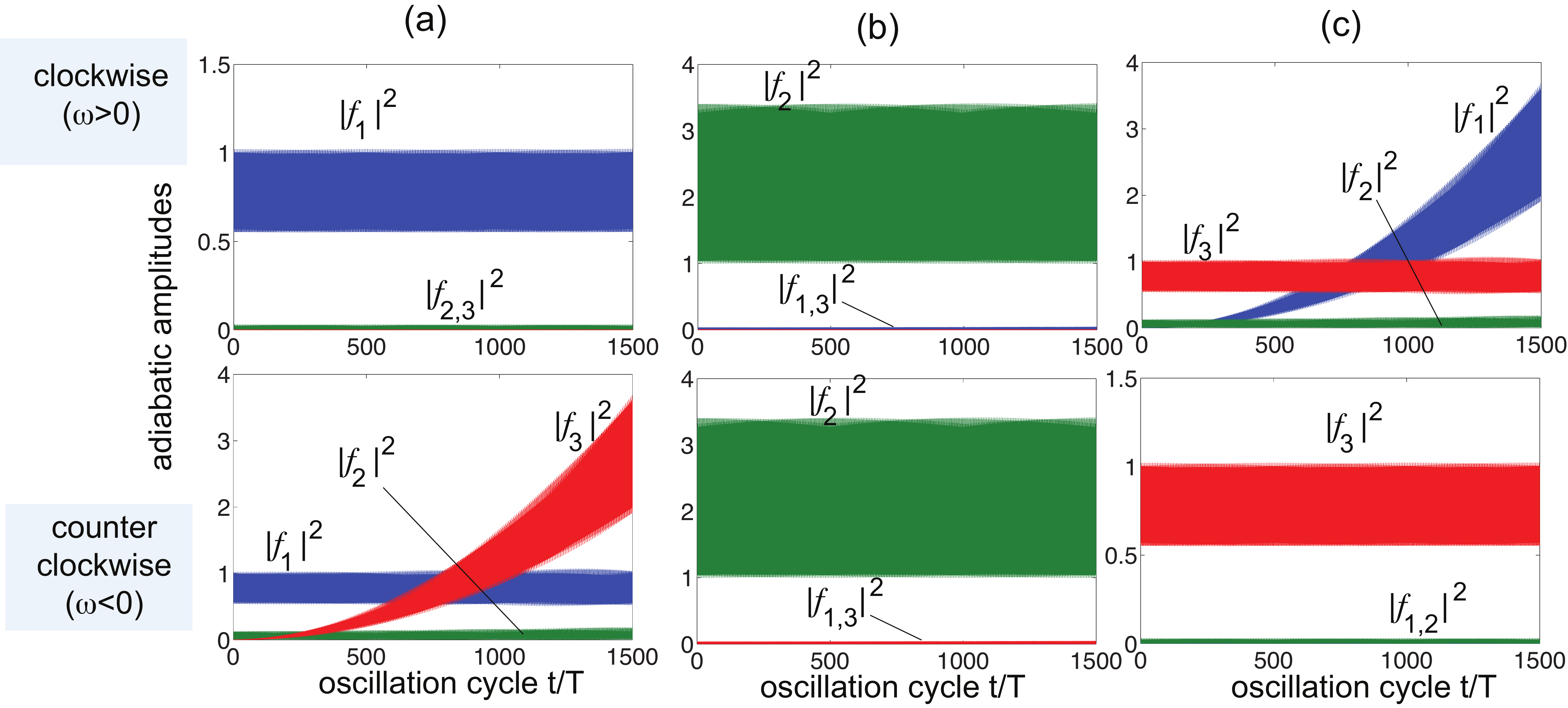}} \caption{
Same as Fig.3, but for parameter values $\Omega=1$, $R_0=0.3$ and $| \omega |=2/7$, corresponding to the odd multiphoton resonance condition $2 \Omega/  \omega=7$.}
\end{figure}

\begin{figure}[htb]
\centerline{\includegraphics[width=15cm]{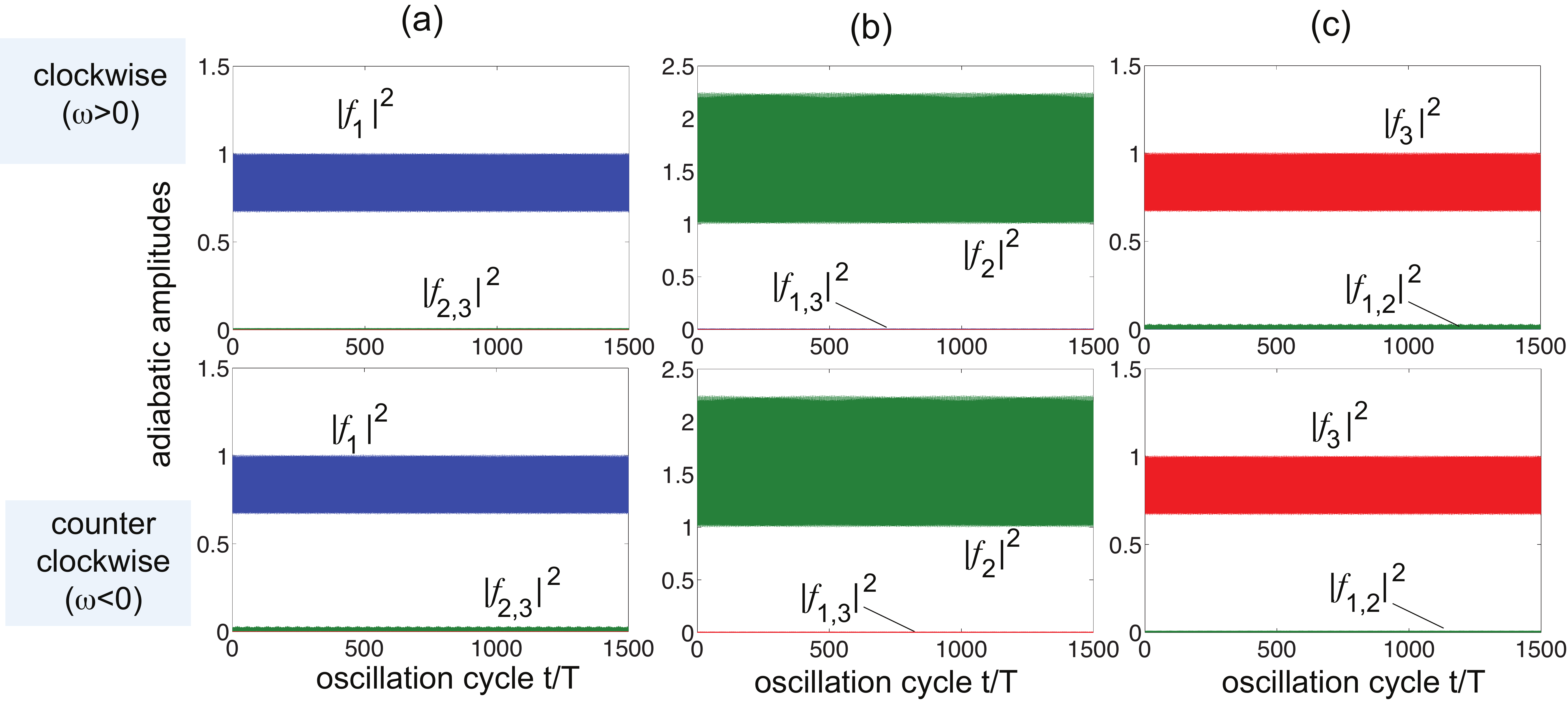}} \caption{
Same as Fig.3, but for $ |\omega|=0.26$. The oscillation frequency does not satisfy a multiphoton resonance condition and nonadiabatic effects are prevented.}
\end{figure}

\section{Conclusions}
In this work we have shown that Floquet exceptional points, corresponding to the coalescence of two (or more) quasi-energies and corresponding Floquet eigenstates of a time-periodic non-Hermitian Hamiltonian, can arise in a wide class of time-periodic systems and can give rise to a chiral dynamics, i.e. selection of a different final state when the system is repeatedly and slowly cycled in opposite directions of parameter space. To conclude, it is worth commenting similarities and differences between the chiral behavior induced by a Floquet EP, studied in this work, and the chirality observed when a static EP is encircled, a well-known phenomenon which has been investigated in several recent works \cite{r18,r25,r36,r37,r38,r40,r43}.  In both cases chirality arises because asymmetry in nonadiabatic effects observed when the circulation direction of the loop is reversed, a typical signature of non-Hermitian dynamics. In case of a Floquet EP nonadiabatic effects are typically observed after many oscillation cycles, they do not require to encircle any static EP, however they arise when a resonance condition is satisfied, which is fully missed when a static EP is slowly cycled. Our results thus provide a different route toward the observation of EP-induced chirality in non-Hermitian systems and are expected to stimulate further theoretical and experimental investigations.

\appendix
\section{Generalized Floquet eigenstates}
In this Appendix we derive the analytical form of the  generalized Floquet eigenstates for the Hamiltonian $\mathcal{H}(t)$ in the presence of a Floquet EP of order $M$. For the sake of definiteness, let us consider the case of a counter-clockwise cycle $(\omega<0$) and let us limit our analysis to a Floquet EP of order $M=2$, corresponding to two eigenvalues $\lambda_{y_1}$ and $\lambda_{y_2}=\lambda_{y1}+G |\omega|$ which become degenerate when folded inside the interval $(-\omega/2, \omega/2)$, i.e. $\tilde{\lambda}_{y_2}=\tilde{\lambda}_{y_1}$. However, the analysis can be readily extended to the case of a Floquet EP of higher order or for a clockwise cycle ($\omega>0$).\\ 
As discussed in the  proof of Theorem 2, the two Floquet eigenstates $\mathbf{f}^{(y_1)}(t)$ and $\mathbf{f}^{(y_2)}(t)$ coalesce into the same state, which is the one emanating from the highest eigenvalue $\lambda_{y_2}$. Such a state, $\mathbf{F}^{(y_2)} (t) \equiv \mathbf{f}^{(y_2)}(t)$, is given by Eqs.(11) and (13) by letting $\lambda_n=\lambda_{y_2}$. The other generalized Floquet eigenstate, $\mathbf{F}^{(y_1)} (t) $, can be obtained looking for a solution to Eq.(1) of the form 
\begin{equation}
\mathbf{a}(t)=\exp(-i \mu t)  \sum_{l=-\infty}^{\infty} \left(  \mathbf{a}^{(l)}  + \gamma t \mathbf{b}^{(l)} \right) \exp(i l \omega t) 
\end{equation}
which includes a secular term that grows linearly in time. In the above equation, the parameter $\gamma$ defines the strength of the secular term and will be determined by a solvability condition. After substitution of the Ansatz (A.1) into Eq.(1) and using Eqs.(9) and (10), the following coupled hierarchical equations for the Fourier amplitudes $\mathbf{a}^{(l)}$ and $\mathbf{b}^{(l)}$ are obtained
\begin{eqnarray}
(\mu-l \omega -\mathcal{H}_0) \mathbf{a}^{(l)} & = & \sum_{k=1}^{\infty} \mathcal{S}^{(k)} \mathbf{a}^{(l-k)}-i \gamma \mathbf{b}^{(l)} \\
(\mu-l \omega -\mathcal{H}_0) \mathbf{b}^{(l)} & = & \sum_{k=1}^{\infty} \mathcal{S}^{(k)} \mathbf{b}^{(l-k)}
\end{eqnarray}
Equation (A.3) can be solved by setting $\mu=\lambda_{y_2}$ and
\begin{equation}
\mathbf{b}^{(l)}= \left\{
\begin{array}{cc}
0 & l <0 \\
\mathbf{w}^{(y_2)} & l=0 \\
 \left( \lambda_{y_2} - l \omega - \mathcal{H}_0 \right)^{-1} \sum_{k=1}^{l} \mathcal{S}^{(k)} \mathbf{b}^{(l-k)}& l \geq 1
\end{array}
\right.
\end{equation}
where $\mathbf{w}^{(y_2)}$ is the eigenvector of $\mathcal{H}_0$ with eigenvalue $\lambda_{y_2}$, i.e. $\mathcal{H}_0 \mathbf{w}^{(y_2)}= \lambda_{y_2} \mathbf{w}^{(y_2)}$. Note that the Fourier terms $\mathbf{b}^{(l)}$ given by Eq.(A.4), which define the secularly growing term in Eq.(A.1), are precisely the ones of the Floquet state $\mathbf{f}^{(y_M)}(t)$, as it should be according to the general Floquet theory of defective quasi-energies (Sec.2). For $ l \neq 0$, Eq.(A.2) is then formally solved by letting
\begin{equation}
\mathbf{a}^{(l)}= \left\{
\begin{array}{cc}
0 & l <-G \\
\mathbf{w}^{(y_1)} & l=-G \\
 \left( \lambda_{y_2} - l \omega - \mathcal{H}_0 \right)^{-1} \sum_{k=1}^{l+G} \mathcal{S}^{(k)} \mathbf{b}^{(l-k)}& -G< l<0 \\
 \left( \lambda_{y_2} - l \omega - \mathcal{H}_0 \right)^{-1} \sum_{k=1}^{l+G} \left( \mathcal{S}^{(k)} \mathbf{b}^{(l-k)} -i \gamma \mathbf{b}^(l) \right)& l \geq 1
\end{array}
\right.
\end{equation}
where $\mathbf{w}^{(y_1)}$ is the eigenvector of $\mathcal{H}_0$ with eigenvalue $\lambda_{y_1}$, i.e. $\mathcal{H}_0 \mathbf{w}^{(y_1)}= \lambda_{y_1} \mathbf{w}^{(y_1)}$. The value $\mathbf{a}^{(0)}$ is obtained as a solution of the inhomogeneous linear system [Eq.(A.2) with $l=0$]
\begin{equation}
\left( \lambda_{y_2}-\mathcal{H}_0 \right) \mathbf{a}_0= \mathbf{d}
\end{equation}
where we have set
\begin{equation}
\mathbf{d} \equiv \sum_{k=1}^G \mathcal{S}^{(k)} \mathbf{a}^{(-k)} - i \gamma \mathbf{w}^{(y_2)}.
\end{equation}
Note that, since $\left( \lambda_{y_2}-\mathcal{H}_0 \right)$ is singular, the inverse of the matrix $\left( \lambda_{y_2}-\mathcal{H}_0 \right)$ is not defined and Eq.(A.6) can be solved provided that the solvability condition 
\begin{equation}
(\mathbf{w}^{(y_2) \dag},\mathbf{d})=0
\end{equation}
is met, where $\mathbf{w}^{(y_2) \dag}$ is the eigenvector of the adjoint matrix $\mathcal{H}_0^{\dag}$ with eigenvalue $\lambda_{y_2}^*=\lambda_{y_2}$. The solvability condition then determines the value of the parameter $\gamma$ entering in the Ansatz (A.1)
\begin{equation}
\gamma=-i \frac{\sum_{k=1}^{G} \left( \mathbf{w}^{(y_2) \dag}, \mathcal{S}^{(k)} \mathbf{a}^{(-k)} \right) } {(\mathbf{w}^{(y_2) \dag},\mathbf{w}^{(y_2)})}.
\end{equation}
Once the solvability condition is satisfied, the solution to Eq.(A.6) can be determined, apart from an arbitrary vector parallel to $\mathbf{w}^{(y_2)}$ which can be chosen to be zero without loss of generality [it would correspond to an additional term proportional to the other Floquet eigenstate $\mathbf{f}^{(y_2)}(t)$].\\
\par

\end{document}